# Surprisingly large inverse spin Hall effect and systematic variation of spin-orbit coupling with *d*-orbital filling in 3*d* transition metals

Chunhui Du[†], Hailong Wang[†], Fengyuan Yang[*], and P. Chris Hammel[*]

Department of Physics, The Ohio State University, Columbus, OH, 43210, USA

[†]These authors made equal contributions to this work

[*]Emails: fyyang@physics.osu.edu; hammel@physics.osu.edu

Abstract

It is generally believed that spin-orbit coupling (SOC) follows $Z^4$ (atomic number) dependence and becomes significant only in heavy elements. Consequently, SOC in 3*d* transition metals should be negligible given their small $Z$. Using dynamic spin pumping of $Y_3Fe_5O_{12}$-based structures, we uncover a systematic evolution of spin Hall angle ($\theta_{SH}$) with *d*-orbital filling in a series of 3*d* metals, reminiscent of behavior observed in 5*d* metals. In particular, Cr and Ni show very large $\theta_{SH}$ (half of that for Pt), indicating that *d*-orbital filling rather than $Z$ plays a dominant role in spin Hall effect (SHE) in 3*d* metals. This result enriches our understanding of SHE and broadens the scope of materials available for exploring the rich phenomena enabled by SOC as well as presenting a guidepost for testing theoretical models of spin-orbit coupling in transition metals.

Spin-orbit coupling is the underlying mechanism for magnetocrystalline anisotropy [1], anomalous Hall effect [2], and more recently, spin Hall effect [3] and topological insulators [4]. It is generally believed that SOC varies as $Z^4$ [5-7], implying that SOC is important only in heavy elements such as $5d$ transition metals, while in lighter elements such as $3d$ transition metals, SOC should be negligibly small. SHE depends on the SOC and the magnitude of $\theta_{SH}$ is a measure of the strength of SOC. Because of the generally accepted $Z^4$ dependence of SOC, measurement of $\theta_{SH}$ has focused on heavy elements, mostly on $5d$ transition metals, by SHE [8] or inverse spin Hall effect (ISHE) [9-15], while $3d$ transition metals have rarely been studied [16, 17].

Ferromagnetic resonance (FMR) spin pumping of pure spin currents from a ferromagnet (FM) into a nonmagnetic (NM) material provides a powerful technique for measurement of $\theta_{SH}$ in a broad range of materials using the ISHE [7, 9]. Here we report a systematic study of the ISHE in a series of $3d$ transition metals using FMR spin pumping from insulating $Y_3Fe_5O_{12}$ (YIG) epitaxial thin films into Ti, V, Cr, Mn, $Fe_{50}Mn_{50}$ (FeMn), $Fe_{20}Ni_{80}$ (Py), Ni, and Cu. Our demonstration of large ISHE signals in our YIG-based structures [7, 15, 18-21] provides unprecedented sensitivity for characterizing the ISHE in $3d$ transition metals which are expected to have weak SOC. Surprisingly, we detect an ISHE voltage ($V_{ISHE}$) exceeding 5 mV in a YIG/Cr(5 nm) bilayer, which is among the highest $V_{ISHE}$ we observed in any materials [7, 15].

We deposit epitaxial YIG films on (111)-oriented $Gd_3Ga_5O_{12}$ (GGG) substrates using off-axis sputtering [7, 15, 22]. A representative $2\theta$-$\omega$ x-ray diffraction (XRD) scan of a 25-nm YIG film shown in Fig. 1(a) demonstrates pure garnet phase and clear Laue oscillations. The x-ray reflectometry (XRR) spectrum of a 40-nm YIG film on GGG in Fig. 1(b) reflects the smooth surface of the YIG film. The atomic force microscopy (AFM) image in the inset to Fig. 1(b) exhibits a smooth surface with a roughness of only 0.106 nm. Figure 1(c) shows the derivative of a FMR absorption spectrum for a 20-nm YIG film taken in a cavity of a Bruker Electron Paramagnetic Resonance (EPR) spectrometer at a radio-frequency (rf) $f$ = 9.65 GHz and an input microwave power $P_{rf}$ = 0.2 mW, which gives a peak-to-peak linewidth ($\Delta H$) of 9.5 Oe. Spin pumping measurements are carried out in the same instrument at room temperature on a series of YIG/metal bilayers (approximate dimensions of 1 mm × 5 mm), in which the thickness of the 3$d$ metal layers is 10 nm. A DC magnetic field $H$ is applied in the $xz$-plane and the ISHE voltage is measured across the ~5-mm long metal layer along the $y$-axis, as illustrated in Fig. 1(d).

Figures 2(a) to 2(f) show $V_{ISHE}$ vs. $H - H_{res}$ spectra ($H_{res}$ is the resonance field of the YIG) of YIG/metal(10 nm) bilayers for Ti, V, Cr, Mn, FeMn, and YIG/Cu(10 nm)/Ni(10 nm) trilayer at two opposite in-plane field orientations $\theta_H$ = 90° and 270° using $P_{rf}$ = 200 mW, which exhibit $V_{ISHE}$ = -24.6 µV, -594 µV, -2.55 mV, -549 µV, -4.65 µV, and 39.4 µV, respectively, at $\theta_H$ = 90°. The negative sign in $V_{ISHE}$ arises from the convention of positive $V_{ISHE}$ for YIG/Pt at $\theta_H$ = 90°. The strong exchange coupling between YIG and Ni induces such substantial additional damping of the YIG that we use YIG/Cu/Ni trilayers to determine $\theta_{SH}$ as reported previously [21]. The 2.5-mV ISHE signal measured in YIG/Cr(10 nm) is exceptionally large and comparable to the values detected in 5$d$ metals Ta, W, and Pt on our YIG films [7, 15]. This suggests unexpectedly large $\theta_{SH}$ and surprisingly strong SOC in Cr. Given the relatively small $Z$ of 3$d$ elements, we explore the potential role of $d$-orbital configuration and antiferromagnetism (AF) [23] in this surprising result. The measurement on antiferromagnetic FeMn is performed to explore the possible role of antiferromagnetism in ISHE. When combined with our previous results on Py [21] and Cu [7], the detailed study of eight 3$d$ transition metals and alloys presented here uncovers unexpected role of $d$-orbital filling in spin Hall physics in this group of light materials.

The mV-level ISHE signal observed in YIG/Cr is quite surprising considering that the SOC

in Cr has been considered negligible due to its small $Z = 24$. Spin Hall angle is a sensitive measure of the strength of SOC and can be calculated from [9, 10, 12, 14],

$$V_{\text{ISHE}} = -e\theta_{\text{SH}} w R \lambda_{\text{SD}} \tanh\left(\frac{t_{\text{NM}}}{2\lambda_{\text{SD}}}\right) g^{\uparrow\downarrow} f P \left(\frac{\gamma h_{\text{rf}}}{4\pi\alpha f}\right)^2, \quad (1)$$

where $e$ is the electron charge, $w$, $R$ and $t_{\text{NM}}$ are the sample width, resistance and thickness, respectively, of the Cr layer, $\lambda_{\text{SD}}$ is the spin diffusion length of Cr, $g^{\uparrow\downarrow}$ is the interfacial spin mixing conductance, $P = 1.21$ is a factor arising from the ellipticity of the magnetization precession [7], $\gamma$ is the gyromagnetic ratio, $h_{\text{rf}} = 0.25$ Oe is the rf field at $P_{\text{rf}} = 200$ mW [7], and $\alpha$ is the Gilbert damping constant of YIG. To calculate $\theta_{\text{SH}}$, we first determine $\lambda_{\text{SD}}$ from the Cr thickness ($t_{\text{Cr}}$) dependence of $V_{\text{ISHE}}$ within 5 nm $\leq t_{\text{Cr}} \leq$ 100 nm [Fig. 2(g)]. The $t_{\text{Cr}}$ dependence of $V_{\text{ISHE}}$ is partially due to the variation in resistivity ($\rho$) of the Cr films as shown in Fig. 2(h), which is similar to the behavior reported previously [24]. The ISHE-induced charge current $I_c = V_{\text{ISHE}}/R$ is proportional to the pure spin current pumped into Cr [7, 15]. Figure 2(i) plots the $t_{\text{Cr}}$ dependence of $V_{\text{ISHE}}/Rw$, from which we obtain $\lambda_{\text{SD}} = 13.3 \pm 2.1$ nm by fitting to $\frac{V_{\text{ISHE}}}{Rw} \propto \lambda_{\text{SD}} \tanh\left(\frac{t_{\text{Cr}}}{2\lambda_{\text{SD}}}\right)$ [25]. The spin mixing conductance can be obtained from the spin-pumping enhancement of damping [9-12],

$$g^{\uparrow\downarrow} = \frac{4\pi M_s t_{\text{YIG}}}{g\mu_B}(\alpha_{\text{YIG/NM}} - \alpha_{\text{YIG}}), \quad (2)$$

where $g$, $\mu_B$, and $t_{\text{YIG}}$ are the Landé $g$ factor, Bohr magneton, and YIG thickness, respectively. We determine the damping constants of a bare YIG film ($\alpha_{\text{YIG}}$) and a YIG/Cr bilayer ($\alpha_{\text{YIG/NM}}$) from the frequency dependencies of the FMR linewidth measured using a microstrip transmission line [Fig. 3(a)]. The linewidth increases linearly with frequency: $\Delta H = \Delta H_{\text{inh}} + \frac{4\pi\alpha f}{\sqrt{3}\gamma}$ [26], where $\Delta H_{\text{inh}}$ is the inhomogeneous broadening. Table I shows the damping enhancement due to spin pumping: $\alpha_{\text{sp}} = \alpha_{\text{YIG/NM}} - \alpha_{\text{YIG}}$, where $\alpha_{\text{YIG/NM}}$ and $\alpha_{\text{YIG}} = (8.7 \pm 0.6) \times 10^{-4}$ are obtained from the least-squares fits in Fig. 3(a). Thus, we calculate $g^{\uparrow\downarrow} = (8.3 \pm 0.7) \times 10^{17}$ m$^{-2}$ for the YIG/Cr interface and $\theta_{SH} = -0.051 \pm 0.005$ for Cr. This surprisingly large $\theta_{SH}$ is half the value of Pt [7].

Using the same approach, we obtain the values of $\theta_{SH}$ for other 3$d$ metals. The spin diffusion lengths of V, Mn, and Ni are determined to be $14.9 \pm 2.4$, $10.7 \pm 1.1$, and $3.2 \pm 0.1$ nm as shown in Figs. 2(j), 2(k), and 2(l), respectively. Considering that V, Cr, and Mn all have similar spin diffusion lengths, and since $\theta_{SH}$ is virtually insensitive to the value of $\lambda_{SD}$ when $\lambda_{SD} \geq t_{NM}$ [due to the term $\lambda_{SD}\tanh(\frac{t_{NM}}{2\lambda_{SD}})$ in Eq. (2)], it is safe to assume $\lambda_{SD}$ of Ti is similar to Cr. The calculated values of $\theta_{SH}$ for Ti and Mn are very small (Table I) while the spin Hall angles for V and Ni are quite large for 3$d$ metals.

To highlight the systematic behavior of $\theta_{SH}$, we plot $\theta_{SH}$ vs. $Z$ in Fig. 3(b) for the eight 3$d$ metals and alloys. We note that V, Cr, and Ni with large $\theta_{SH}$ sit directly above Ta, W, and Pt in the periodic table, respectively, which exhibit some of the largest $\theta_{SH}$. This suggests that the $d$-electron configuration of the transition metals plays a very important role in SHE, consistent with the prediction of Tanaka $et~al.$ [27] who illuminated the role of the total number of 4$d$ (5$d$) and 5$s$ (6$s$) electrons in the SHE in the 4$d$ (5$d$) transition metals. To understand the role of $d$-electrons, we list in Table I the total number of 3$d$ and 4$s$ electrons, $n_{3d+4s}$. We note that among these, $\theta_{SH}$ varies significantly both in sign and magnitude: $\theta_{SH}$ is negative from Ti ($n_{3d+4s} = 4$) to FeMn ($n_{3d+4s} = 7.5$) and changes to positive for Py ($n_{3d+4s} = 9.6$), Ni ($n_{3d+4s} = 10$) and Cu ($n_{3d+4s} = 11$) while its magnitude reaches maximum at Cr ($n_{3d+4s} = 6$) and Ni ($n_{3d+4s} = 10$). The sign change in $\theta_{SH}$ mimics the trend observed in 5$d$ metals [7, 13, 27, 28], while the magnitude of $\theta_{SH}$ spans a range of almost three orders of magnitude. From Fig. 3(b) and our previous result on 5$d$ metals [7], we can gain insights into the underlying mechanisms responsible for the SHE and SOC in transition metals.

There are three mechanisms that could be responsible for the spin Hall effect in transition metals: 1) atomic number, 2) $d$-electron count, and 3) magnetic ordering; we address these separately below. First, while the atomic number may play a role in SHE in 3$d$ metals, it is not a dominant factor: for example, between Cr and W which belong to the same VIB transition metal group, the $Z^4$ dependence predicts a difference of 90 times in their SOC strengths and $\theta_{SH}$, while our experimental values show a factor of 2.7 in $\theta_{SH}$ between the two elements.

Secondly, we can also rule out magnetic ordering in the 3$d$ metals as the dominant factor. While Cr and Ni exhibit large $\theta_{SH}$, they also possess magnetic ordering: Cr is an antiferromagnet [23] and Ni is a ferromagnet. To probe the role of AF ordering in ISHE in 3$d$ metals, we compare the spin Hall angles of Cr and FeMn, a robust antiferromagnet. The spin Hall angle of Cr is 689 times larger than that of FeMn (Table I). The dramatic difference in the two 3$d$ AF metals suggests that the surprisingly large $\theta_{SH}$ in Cr does not arise from its AF order [29]. The very small $\theta_{SH}$ of FeMn also agrees with the theoretical prediction for 4$d$ and 5$d$ metals [27] in that at $n_{3d+4s} \approx 7.5$, the spin Hall conductivity (SHC) crosses zero.

For FM metal Ni, we consider the two elements directly below Ni in the periodic table, Pd and Pt. Tanaka *et al.* [27] calculate that, when $\rho$ is in an appropriate range, the SHC of Pd is ~70% of that for Pt, much larger than the 11% predicted from the $Z^4$ dependence. If we similarly assume a 70% ratio in SHC for Ni relative to Pd, we would conclude that the SHC for Ni is 49% that of Pt: very close to our experimentally measured ratio. This is without considering the FM ordering in Ni. A theoretical calculation of the SHE in 3$d$ metals is needed to verify this explanation. Thus, the surprisingly large values and significant variation in $\theta_{SH}$ of 3$d$ metals arise mainly from the $d$-electron configuration, indicating its dominant role in spin Hall physics [7, 13, 27, 28].

Taken together, our results in 3$d$ and 5$d$ [7] metals reveal a surprising feature of ISHE: the effects of atomic number and *d*-orbital filling are additive—not multiplicative—indicating they operate independently, and each of these mechanisms can be of comparable importance. This means that if either contribution (*Z* or *d*) is large, the spin Hall effect is large, not that if either one is small, the spin Hall effect is small. For example, the $Z^4$ dependence is clearly dominant in the Cu, Ag, and Au series [7] whose filled *d*-shells have zero orbital moment and do not contribute to the ISHE; while for transition metals with partially filled *d*-orbitals, the *d*-orbital contribution to the ISHE is dominant, as demonstrated by the variation of sign and magnitude of spin Hall angles in both 3*d* and 5*d* transition metals.

Furthermore, we confirm the influence of Cr antiferromagnetism on the static and dynamic magnetization of YIG. Cr is an incommensurate AF with a Néel temperature of 311 K in the bulk [23]. In Cr thin films, the AF ordering temperature is reduced to below room temperature. The static or dynamic AF ordered spins in Cr are expected to couple to the YIG magnetization via interfacial exchange interaction [30], resulting in possible exchange bias and enhanced coercivity ($H_c$) [31]. The room temperature, in-plane magnetic hysteresis loops for a 20-nm YIG film and YIG/Cr($t_{Cr}$) bilayers with $t_{Cr}$ = 10, 35, 50, and 100 nm shown in Fig. 4 demonstrate that this is in fact the case. The bare YIG film exhibits a square hysteresis loop with a very small $H_c$ = 0.35 Oe and a very sharp magnetic switching. At $t_{Cr}$ = 10 nm, $H_c$ only increases slightly to 0.52 Oe, suggesting that at 10 nm, the correlation of Cr spins is fairly weak. As $t_{Cr}$ increases, $H_c$ continuously rises and reaches 1.73 Oe, indicating strengthening AF correlation with increasing $t_{Cr}$. This observation is further verified by the magnetic damping enhancement shown in the inset to Fig. 4, where the YIG/Cr(100 nm) exhibits a much larger damping constant than the YIG/Cr(10 nm), corroborating stronger AF correlation in thicker Cr films. As a comparison, both the 10-nm and 100-nm vanadium films induce similar damping in YIG due to its nonmagnetic nature.

In conclusion, we observe surprisingly large, mV-level ISHE voltages in YIG/Cr bilayers and robust spin pumping signals in other 3$d$ metals. By measuring ISHE voltages and damping enhancement, we determine the spin Hall angles of eight 3$d$ metals, which reveal unexpected systematic behavior involving both sign change and dramatic variation in magnitude, implying the dominant role of $d$-electron configuration in SHE of 3$d$ metals. Theoretical calculations similar to those performed for 4$d$ and 5$d$ transition metals [27] are needed for thorough understanding of the underlying mechanisms responsible for the observed large SHE in 3$d$ transition metals.

This work was primarily supported by the U.S. Department of Energy (DOE), Office of Science, Basic Energy Sciences, under Award# DE-FG02-03ER46054 (FMR and spin pumping characterization) and Award# DE-SC0001304 (sample synthesis and magnetic characterization). This work was supported in part by the Center for Emergent Materials, an NSF-funded MRSEC under award # DMR-0820414 (structural characterization). Partial support was provided by Lake Shore Cryogenics Inc. and the NanoSystems Laboratory at the Ohio State University.

**Table I**. Total number of 3d and 4s electrons ($n_{3d+4s}$), ISHE voltages at $f$ = 9.65 GHz and $P_{rf}$ = 200 mW, Gilbert damping enhancement due to spin pumping $\alpha_{sp} = \alpha_{YIG/NM} - \alpha_{YIG}$ ($\alpha_{YIG} = 8.7 \pm 0.6 \times 10^{-4}$) and the calculated interfacial spin mixing conductance, electrical resistivity, spin diffusion length, and spin Hall angle for each metal (alloy).

| Bilayer/Trilayer | $n_{3d+4s}$ | $\|V_{ISHE}\|$ | $\alpha_{sp}$ | $g^{\uparrow\downarrow}$ (m$^{-2}$) | $\rho$ ($\Omega$ m) | $\lambda_{SD}$ (nm) | $\theta_{SH}$ |
|---|---|---|---|---|---|---|---|
| YIG/**Ti** | 4 | 24.6 µV | $(1.8 \pm 0.1) \times 10^{-3}$ | $(3.5 \pm 0.3) \times 10^{18}$ | $3.0 \times 10^{-6}$ | ~13.3 | $-(3.6 \pm 0.4) \times 10^{-4}$ |
| YIG/**V** | 5 | 594 µV | $(1.6 \pm 0.1) \times 10^{-3}$ | $(3.1 \pm 0.3) \times 10^{18}$ | $2.9 \times 10^{-6}$ | 14.9 | $-(1.0 \pm 0.1) \times 10^{-2}$ |
| YIG/**Cr** | 6 | 2.55 mV | $(4.3 \pm 0.3) \times 10^{-4}$ | $(8.3 \pm 0.7) \times 10^{17}$ | $8.3 \times 10^{-6}$ | 13.3 | $-(5.1 \pm 0.5) \times 10^{-2}$ |
| YIG/**Mn** | 7 | 549 µV | $(2.3 \pm 0.2) \times 10^{-3}$ | $(4.5 \pm 0.4) \times 10^{18}$ | $9.8 \times 10^{-6}$ | 10.7 | $-(1.9 \pm 0.1) \times 10^{-3}$ |
| YIG/**FeMn** | 7.5 | 4.65 µV | $(2.5 \pm 0.2) \times 10^{-3}$ | $(4.9 \pm 0.4) \times 10^{18}$ | $2.8 \times 10^{-6}$ | 3.8 [32] | $-(7.4 \pm 0.8) \times 10^{-5}$ |
| YIG/Cu/**Py** | 9.6 | 23.7 µV | $(3.3 \pm 0.3) \times 10^{-3}$ | $(6.3 \pm 0.5) \times 10^{18}$ | | 1.7 | $(2.0 \pm 0.5) \times 10^{-2}$ |
| YIG/Cu/**Ni** | 10 | 39.4 µV | $(1.0 \pm 0.1) \times 10^{-3}$ | $(2.0 \pm 0.2) \times 10^{18}$ | | 3.2 | $(4.9 \pm 0.5) \times 10^{-2}$ |
| YIG/**Cu** | 11 | 0.99 µV | $(8.1 \pm 0.6) \times 10^{-4}$ | $(1.6 \pm 0.1) \times 10^{18}$ | $6.3 \times 10^{-8}$ | 500 | $(3.2 \pm 0.3) \times 10^{-3}$ |

**Figure captions:**

**Figure 1.** (a) Semi-log $2\theta$-$\omega$ XRD scan of a 25-nm thick YIG film near the YIG (444) peak, which exhibits clear Laue oscillations. (b) X-ray reflectometry spectrum of a YIG(40 nm) film on GGG. Inset: AFM image of a 25-nm YIG film with a roughness of 0.106 nm. (c) A representative room-temperature FMR derivative spectrum of a YIG film with an in-plane field at $P_{rf}$ = 0.2 mW, which gives a peak-to-peak linewidth of 9.5 Oe. (d) Schematic of experimental setup for ISHE measurements.

**Figure 2.** $V_{ISHE}$ vs. $H - H_{res}$ spectra of (a) YIG/Ti, (b) YIG/V, (c) YIG/Cr, (d) YIG/Mn, (e) YIG/Fe$_{50}$Mn$_{50}$ bilayers and (f) YIG/Cu/Ni trilayer at $\theta_H$ = 90°(red) and 270° (blue) using $P_{rf}$ = 200 mW. Cr thickness dependence of (g) ISHE voltage, (h) resistivity, and (i) ISHE-induced charge current ($V_{ISHE}/R$) normalized by sample width $w$ of YIG/Cr($t_{Cr}$) bilayers. A spin diffusion length of $\lambda_{SD}$ = 13.3 $\pm$ 2.1 nm is obtained from (i). (j) V, (k) Mn, and (l) Ni thickness dependencies of ISHE-induced charge currents normalized by $w$ of the YIG/V($t_V$), YIG/Mn($t_{Mn}$), and YIG/Ni($t_{Ni}$) bilayers give $\lambda_{SD}$ = 14.9 $\pm$ 2.4, 10.7 $\pm$ 1.1, and 3.2 $\pm$ 0.1 nm for V, Mn, and Ni, respectively.

**Figure 3.** (a) Frequency dependencies of FMR linewidth of a bare YIG film, five YIG/metal bilayers, and a YIG/Cu/Ni trilayer. (b) Atomic number ($Z$) dependence of the calculated spin Hall angles ($\theta_{SH}$) of 3$d$ metals and alloys shows a surprisingly large variation of $\theta_{SH}$ with $n_{3d+4s}$.

**Figure 4.** Influence of film thickness on Cr antiferromagnetism: room temperature magnetic hysteresis loops of a single YIG(20 nm) film and YIG/Cr bilayers with Cr thicknesses of 10, 35, 50, and 100 nm, which give coercivities of 0.35, 0.52, 0.74, 1.26, and 1.73 Oe, respectively. The inset shows the frequency dependencies of FMR linewidth of YIG/Cr(10 nm) and YIG/Cr(100 nm).

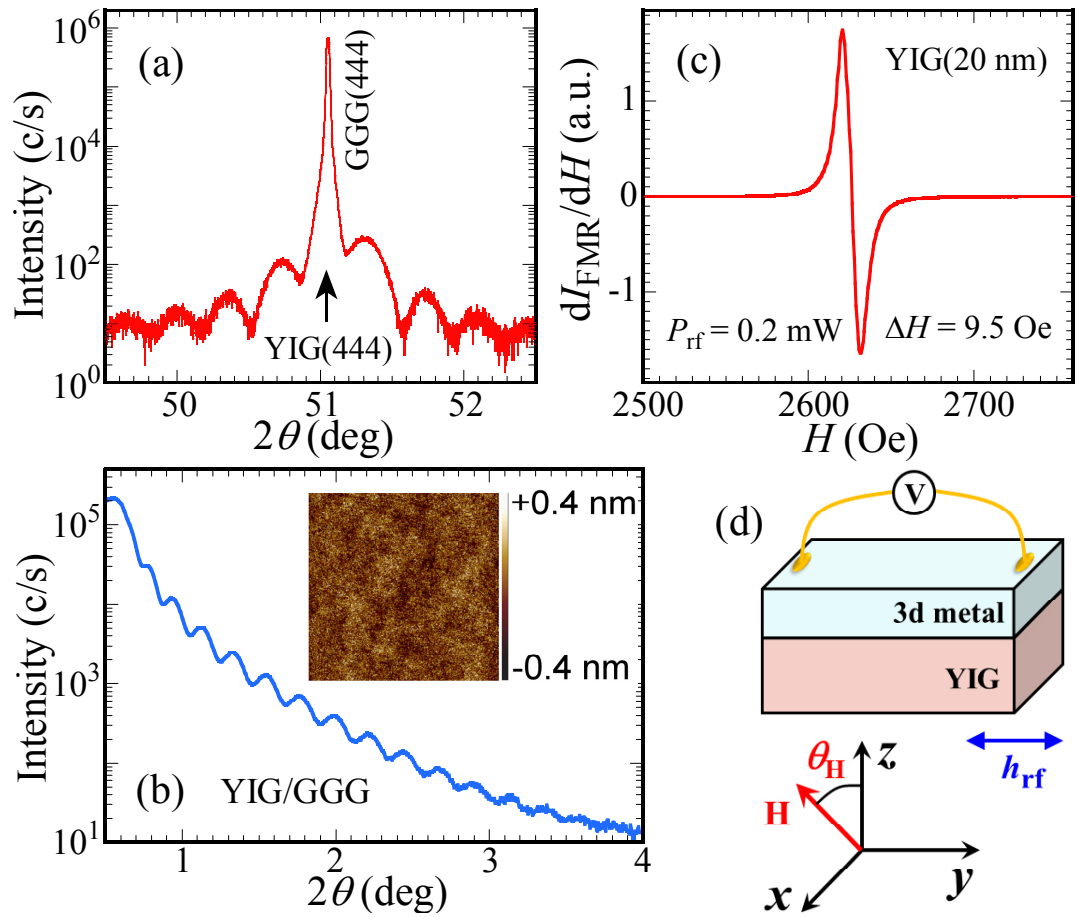

**Figure 1**

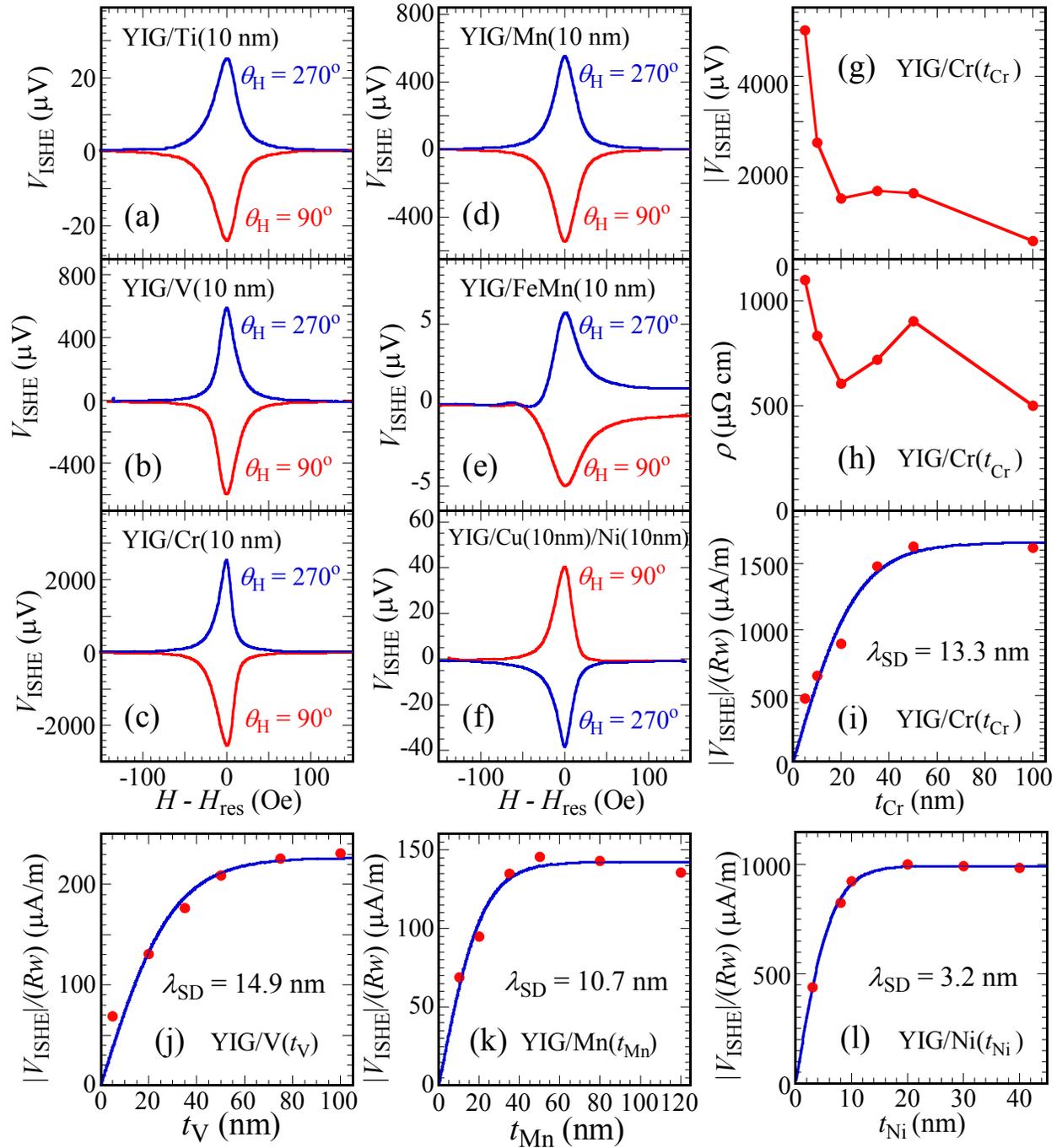

**Figure 2**

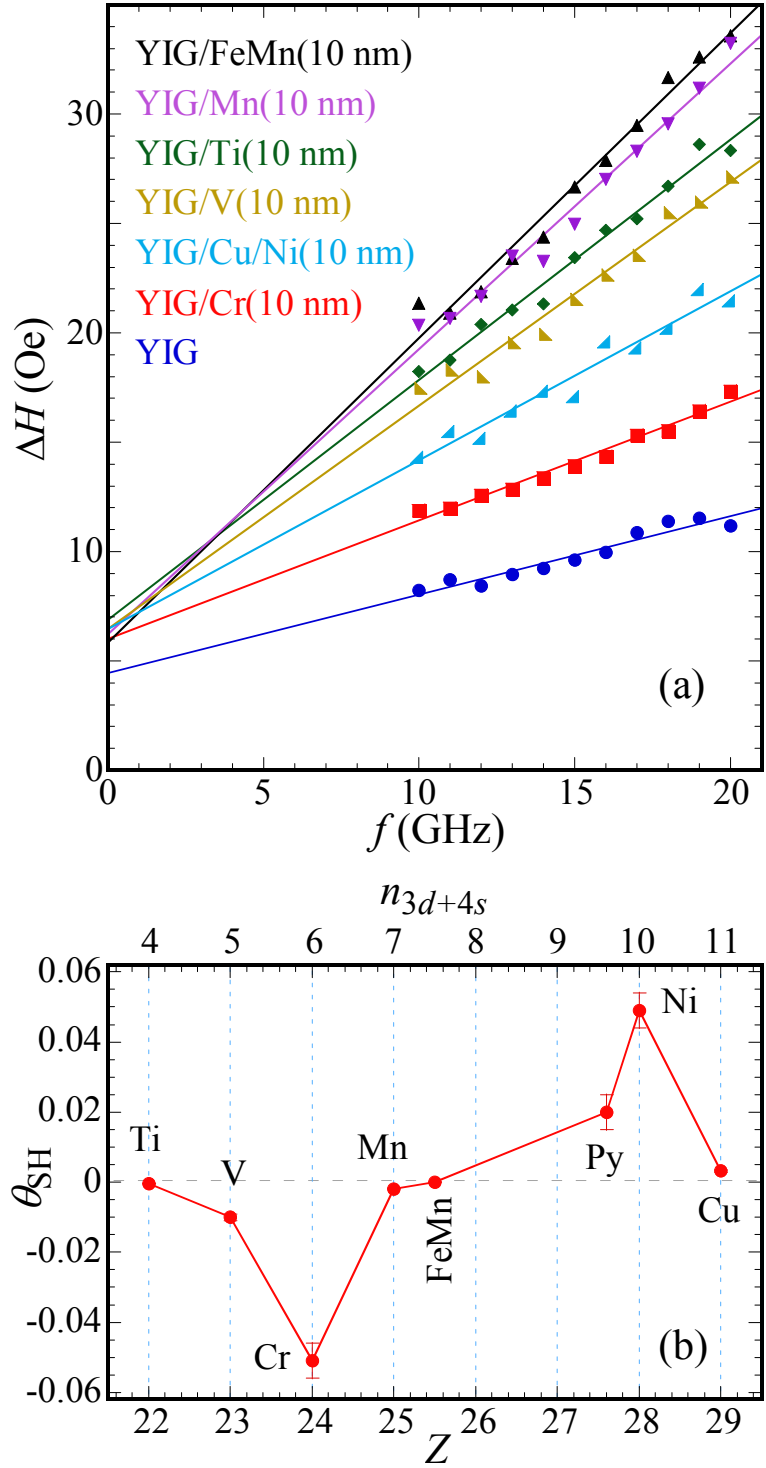

**Figure 3**

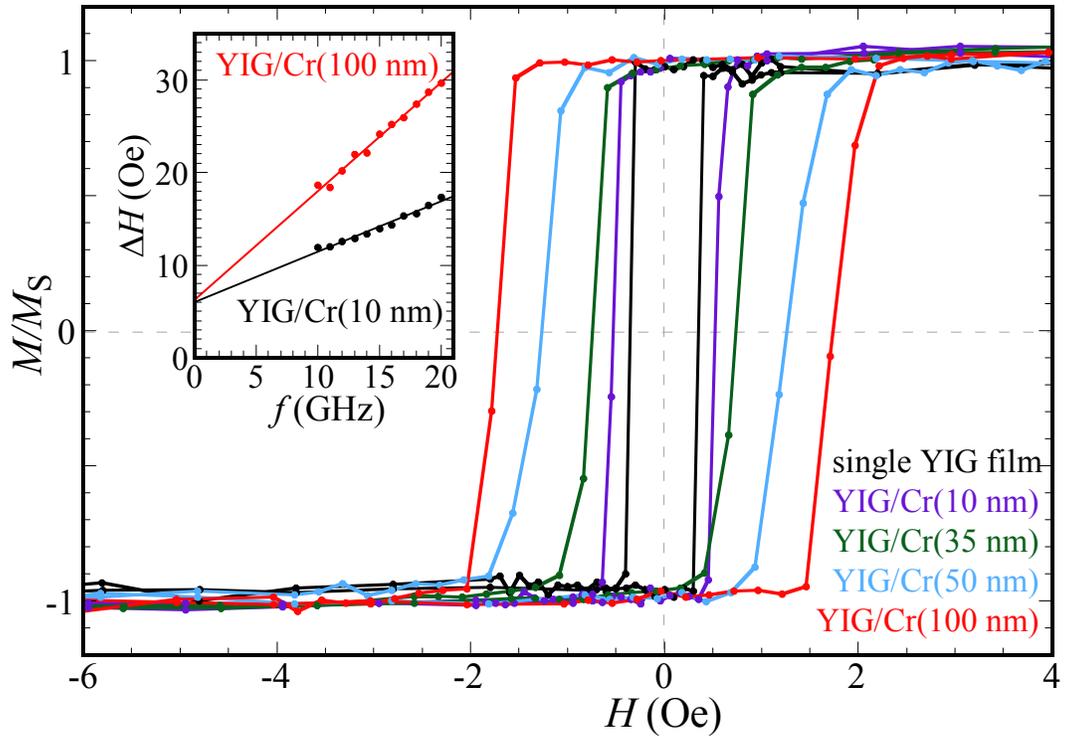

**Figure 4**